\documentclass[aps, prl, aps10pt4-1, twocolumn, apsshowpacs, floatfix, reprint] {revtex4-1}

\usepackage[utf8]{inputenc}	% Prise en compte des accents
\usepackage{graphicx}		% Pour inclure les fichiers de figures (images)
\usepackage{dcolumn}		% Aligne les colonnes d'un tableau sur le nombre de decimal
\usepackage{bm}			% Mettre les symbôle mathematique en gras
\usepackage{amsmath,amssymb}	% Pour les different style mathématique (et pouvoir encadrer une formule avec \boxed)
\usepackage[pdftex,
  bookmarksopen=false,
  colorlinks=true,
  pagebackref=false,			% Chemin retour du lien possible
  citecolor=blue,			% Couleur des liens de bibligraphie
  linkcolor=black,			% Couleur des liens de SOMMAIRE
  colorlinks=true,			% Colorie le texte et ne met pas d'encadrement
  plainpages=false]{hyperref}		% Ajoute de liens hypertext
\usepackage{subscript}
\usepackage{textcomp}		% Pour le symbole °
\usepackage{color}		% Mettre du texte en couleur
\usepackage{lipsum}

\usepackage{epsfig}		% Placer des figure en eps
\usepackage{subfigure}		% Placer plusieurs graphiques l'un sous l'autre

\newcommand{\notedebase}[1]	% Permet d'insérer des notes |e| 2 fois 2 lignes en caractère plus petit pour garder certaines idées
           {\raggedright \footnotesize 
	     \vskip-0.5 \baselineskip \rule[-1.4mm]{\linewidth}{0.5pt} \rule[1.4mm]{\linewidth}{0.5pt} \\ \vskip-0.5\baselineskip #1\\
             \vskip-\baselineskip \rule[-1.4mm]{\linewidth}{0.5pt} \rule[1.4mm]{\linewidth}{0.5pt} \\}

\begin{document}
\title{Evidence of non collisional femtosecond laser electron heating in dielectric materials}
\author{G. Duchateau\textsuperscript{*}, B. Chimier, S. Coudert, E. Smetanina, L. Barilleau, N. Fedorov, H. Jouin, G. Geoffroy, P. Martin, and V.T. Tikhonchuk}
\affiliation{Université de Bordeaux-CNRS-CEA, Centre Lasers Intenses et Applications, UMR 5107, 351 Cours de la Libération, 33405 Talence, France}

\email{guillaume.duchateau@u-bordeaux.fr}

\date{\today}

\begin{abstract}
Electron dynamics in the bulk of large band gap dielectric crystals induced by intense femtosecond laser pulses at 800 nm is studied. With laser intensities under the ablation threshold (a few 10 TW/cm\textsuperscript{2}), electrons with unexpected energies in excess of 40-50 eV are observed by using the photoemission spectroscopy. A theoretical approach based on the Boltzmann kinetic equation including state-of-the-art modeling for various particles interactions is developed to interpret these experimental observations. A direct comparison shows that both electron heating in the bulk and a further laser field acceleration after ejection from the material contribute equivalently to the final electron energy gain. The electron heating in the bulk is shown to be significantly driven by a non-collisional process, i.e. direct multiphoton transitions between sub-bands of the conduction band. This work also sheds light on the contribution of the standard electron excitation/relaxation collisional processes, providing a new baseline to study the electron dynamics in dielectric materials and associated applications as laser material structuring.
%	Experimental energetic photoelectron (20 to 60 eV) emitted from a  $\alpha$-quartz dielectric crystal irradiated by intense femtosecond laser pulses are a not expected Fermi-Dirac shapes. These observations are interpreted using a model based on the resolution of Boltzmann equation coupling with fundamental equation of the classical dynamics. We demonstrate that these spectra are due to heating and relaxation processes in the bulk coupling with laser-driven acceleration occuring after the electron ejection from the target surface.
%	Experiment in dielectric photoemission spectroscopy show that energetic photoelectron (20 to 60 eV) emitted from $\alpha$-quartz irradiated by intense femtosecond laser pulses are a specific Fermi-Dirac shapes. The underlying physic is that the electrons are heated and relax in the bulk material by several processes before escaped from the target surface. The ejected electrons are accelerated in the vacuum with laser-driven acceleration which give the final stucture of the spectra. These observations are interpreted using a complete physical model based on the numerical resolution of Boltzmann equation including phonon and interband heating processes in the material, coupling with fundamental equation of the classical dynamics in the vacuum. This model permits to understand electron dynamics at 800 nm and is succesfully compared with experimental spectra.  
\end{abstract}

\pacs{PACS}	% Schéma de classification hierarchique : Physics and Astronomy Classification Scheme (PACS)
	
\maketitle

%%%%%%%%%%%%%%%%%%%%%%%%%%%%%%%%%%%%%%%%%%%%%%%%%%%%%%%%%%%%%%%%%%%%%%%%%%%%%%%%%%%%%%%%%%%%%%%%%
%%%%%%%%%%%%%%%%%%%%%%%%%%%%%%%%%%%%%%%%%%%%%%%%%%%%%%%%%%%%%%%%%%%%%%%%%%%%%%%%%%%%%%%%%%%%%%%%%
Developments of laser facilities delivering ultrashort and intense laser pulses with photon energy in the eV range have motivated studies in laser-solid interactions including metals \cite{Rethfeldb, PhysRevB.87.035414}, semiconductors \cite{PhysRevB.54.4660} and dielectrics \cite{Kaiser, Shchelebanov}. Focusing of a femtosecond laser pulse in a transparent dielectric material may induce modifications beneath the surface, which can be tailored to produce permanent three dimensional localized structural changes \cite{Glezer96, Gamaly, Juodkazis, Mazur, PhysRevLett.112.033901}. These nano-structurations depend on the amount of laser energy deposited in the irradiated volume. A control of the amount and spatial shape of the deposited laser energy opens the way to a large variety of applications going from photonics, bulk microelectronics, nano-fluidics, to medicine \cite{Menzel}. Together with advanced experimental setups, such a control can be achieved by an in-depth modeling description of the physical processes at play. An accurate prediction of the laser energy deposition may further support the development of these applications and improve the knowledge of the fundamental laws governing the laser-solid interaction. So there is a strong need to accurately describe the electron dynamics in dielectric materials irradiated by femtosecond laser pulses with intensities ranging from a few TW/cm$^2$ to the ablation threshold.

The admitted picture for the laser energy deposition into the dielectric material is as follows. The laser energy is first absorbed by electrons through the processes of both ionization and excitation/relaxation in the conduction band (CB). During the second stage, the absorbed laser energy is redistributed between the excited carriers which may reach higher energies while they undergo collisions with phonons, ions, and other electrons in the presence of the laser field. These processes eventually lead to the energy transfer to the lattice. This is a collisional picture theoretically described either by the Drude model, multiple rate equations \cite{Rethfeld}, or the kinetic Boltzmann equation \cite{Kaiser, Shchelebanov, Barilleau}. However, photoemission spectroscopy experiments \cite{NanoLett.13.674, Yatsenko2005, Bachau2010, Belsky2004a}, providing the energy distribution of electrons ejected from the sample surface have shown electrons with energies in excess of tens of eV for laser intensities below the breakdown threshold \cite{Belsky2004b, Belsky2004a}, of which collisional heating is not able to account for. It has been suggested that it is due to direct multiphoton transitions between sub-bands of the CB, hereafter referred to as the interband process \cite{Yatsenko2005, Bachau2010}. The previous observations have been obtained for various dielectric materials including CsI, diamond, CeF$_3$, sapphire, and SiO$_2$ \cite{Belsky2004b, Belsky2004a}, highlighting an universal behavior. A question then arises on the importance of the interband process relative to collision-assisted electron transitions, and on its contribution to the laser energy deposition in dielectric materials (which is related to the electron energy distribution).

Two main classes of models can address this question. (i) Collisional models as solving the state-of-the-art quantum Boltzmann equation, including all possible electronic excitation and relaxation processes, provide the electron energy distribution \cite{Kaiser, Rethfeld, Shchelebanov}. But the interband process has never been included except in \cite{Barilleau}. (ii) Non collisional models based on a resolution of the time-dependent Schrödinger equation support the photoemission observation but the collisions are not included \cite{Yatsenko2005, Bachau2010}. Despite these studies, a \textit{direct} comparison between theoretical and experimental electron energy distributions has never been performed while this is an indispensable step for validating any model \cite{Kaiser, Rethfeld, Shchelebanov, Yatsenko2005}, leaving serious interrogations regarding the (non-equilibrium) electron dynamics in the conduction band of dielectric materials. 
\begin{figure}[t]
  \includegraphics[width=12cm,trim = 0cm 0cm -8cm 0cm,clip=]{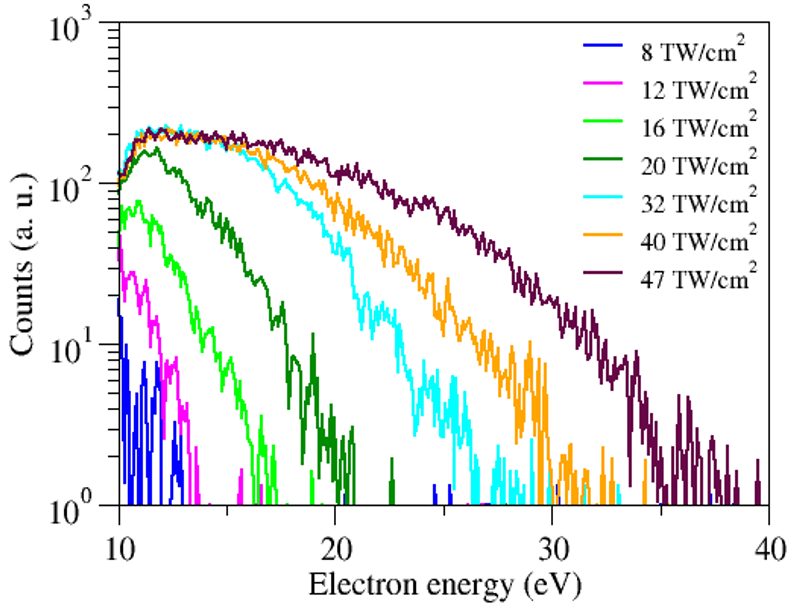}
  \caption{Experimental photoelectron energy distribution for quartz. Only the region of highest electron energies is shown, where dynamics is only due to intrinsic processes modeled in the present study (see text for more details).}
  \label{fig:exp_signal}
\end{figure}

In the present work, the electron dynamics is studied both experimentally and theoretically for $\alpha$-quartz (cristalline SiO$_2$) which is a representative example of large band gap dielectrics. The observed photoelectron energy distributions are \textit{directly} compared, within the challenging accuracy-demanding linear scale in the present context, to a model including the electron dynamics both inside the bulk and after ejection. It is described by a Boltzmann equation including all possible bulk excitation/relaxation processes coupled to a subsequent laser driven field acceleration (LDFA) of electrons after their ejection \cite{NPHYS1983}. The model allows to predict photoelectron spectra which are in a good agreement with experimental results for various laser intensities. The non collisional direct multiphoton transitions between sub-bands of the CB make a significant heating of electrons in dielectric materials in contrast to the widely used assumption of a dominant role of collisional processes including phonon-assisted photon absorption and inverse Bremsstrahlung. Our model accounts in particular for acceleration of electrons to energies in excess of 40 eV below the ablation threshold.

The experiment is carried out on the Aurore Ti:saphire laser facility \cite{Appl.Phys.B.70.S165}. A 1 mm thick $\alpha$-quartz target is irradiated by linearly P-polarized pulses at the wavelength of $\lambda=$ 800 nm, with 70 fs duration (full width at half maximum, FWHM) at 1 kHz repetition rate. The incident angle is 45° and a 30 mm lens produced a Gaussian intensity distribution in a 21 µm spot size (FWHM). The experiment is conducted in a vacuum chamber at a pressure of $10^{-9}$ Torr, and the sample is heated homogeneously to a temperature of 800 K to maximize the photoemission yield, by decreasing the surface charge. The photoelectrons emitted from the surface are collected by a hemispherical analyser (CLAM IV VG Microtech) with 9 channeltrons operating in ultraviolet photo-electron spectroscopy mode. The axis of the detector is perpendicular to the sample surface.

Figure \ref{fig:exp_signal} shows the experimental photo-emission spectra for quartz obtained with intensities ranging from 8 to 47 TW/cm\textsuperscript{2}. In general, such spectra exhibit a main peak, for an electron energy of a few eV, which almost does not evolve with respect to the intensity. This peak corresponds to secondary electrons which properties depend on the surface state. Since we are interested in intrinsic processes corresponding to high enough energies, the low energy region is not shown. Above roughly 11 eV, the signal exhibits a smooth decrease up to a maximal energy, $E_{max}$, for which at least one count is measured. $E_{max}$ increases with respect to the intensity and reaches roughly 40 eV for the highest intensity used below the ablation threshold. Similar behaviors in terms of high energies and distribution shape have been obtained for other large bandgap materials as sapphire, CsI, diamond, and CeF$_3$ \cite{Belsky2004b, Belsky2004a}.

The electron dynamics in the bulk is described by a Boltzmann kinetic equation~\cite{Rethfeld, Barilleau, Shchelebanov}. Electrons are ejected from a nanometer-size layer beneath the target surface where the laser electric field can be considered as a constant. Indeed, the field amplitude adapts to the dielectric material property on a lengthscale $~ v_b / \omega_{ve}$ where $v_b$ is the velocity of bound valence electrons and $\omega_{ve}$ is their plasma frequency. The order of magnitude of these quantities is $3.\times 10^{6} m.s^{-1}$ and $10^{16} s^{-1}$, respectively, leading to $v_b / \omega_{ve} \simeq 3 \AA$. The laser intensity is thus relatively constant a few nanometers beneath the surface, implying no spatial dependence in the electron distribution, $f(\vec{k},t)$, where $\vec{k}$ is the momentum and the electron energy is $E_k = \hbar^2 {\vec{k}}^2 / 2m_e$. The temporal evolution of $f$ in the bulk is then given by:
%	To understand this observation, a theorical model is developed based on the standard three-step approach for electron photoemission~\cite{PhysRev.136.A1030, PhysRev.136.A1044, J.Phys.Chem.Solids.59.527}. Firstly, conduction electrons are produced by excitation of the solid target by the laser field. Secondly, electrons are transported through the solid and can undergo various collisions. Thirdly, electron coming from the bulk with an energy greater than the work function can escaped from the target surface.
%	Those first and second steps are modelized by solving the Boltzmann equation. In our approach, the laser field is considered homogeneous in space so that transport processes are neglected. In addition, the assumption of localized interaction is used, and the contribution of external forces is neglected. Finally, only the temporal evolution of the non-equilibrium electron energy distribution function $f(\vec k,t)$  in the CB of the solid is considered, which leads to numerically solve~\cite{PhysRevB.61.11437, Appl.Phys.A.110.579}:
\begin{equation}
  \dfrac{\partial}{\partial t} f(\vec k, t)  =  \left. \dfrac{\partial f(\vec k, t)}{\partial t} \right|_{_{\text{ioniz}}}
  + \left. \dfrac{\partial f(\vec k, t)}{\partial t} \right|_{_{\text{relax}}} + \left. \dfrac{\partial f(\vec k, t)}{\partial t} \right|_{_{\text{heat}}} \text{,}
  \label{eq:Boltzmann}
\end{equation}
where the three collision integrals in the right hand side describe the ionization, the relaxation, and the laser excitation of conduction electrons, respectively. The electron distribution is assumed to be isotropic since it is due to electron collisions with acoustic phonons which characteristic timescale is 10 fs \cite{Daguzan}.

The ionization processes consist of both the photo-ionization, which is evaluated through the complete Keldysh expression~\cite{Keldysh}, and the impact ionization described as in \cite{Kaiser}. The relaxation processes are related to electron-electron (e-e)~\cite{Kaiser} and electron-phonon (e-ph)~\cite{Kaiser, Daguzan} collisions, which induce the energy exchange of electrons between themselves and to the lattice, respectively. These collision integrals are calculated with the Fermi's golden rule. The electron recombination is also included with a characteristic time of 150 fs \cite{Martin} which is assumed not to depend on $E_k$. The energy distribution of phonons is assumed not to evolve during this short interaction time and is set to the equilibrium Bose-Einstein distribution with a lattice temperature set to 800 K.
\begin{figure}[t]
  \includegraphics[width = 9cm, trim = 0cm 0cm 0cm 3cm, clip] {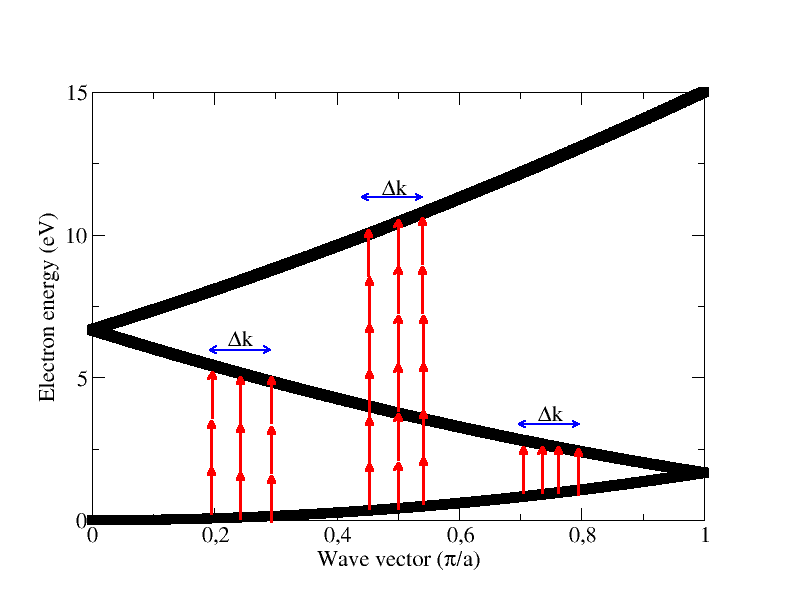}
  \caption{Illustration of the structure of the conduction band in the first Brillouin zone as described in the multiple parabolic band model. The bands exhibit an energy bandwidth of $h\nu_c$ of the order of 1 eV due to the collisional broadening. An illustration of possible multiphoton transitions is depicted by the red arrows. Due to the broadening, the transitions take place over a wave vector region $\Delta k$ \cite{suppMat}.}
  \label{fig:interband}
\end{figure}

Two main processes are included for the laser-induced excitation of conduction electrons. First, the electrons can absorb or emit simultaneously several photons during a collision with phonons or ions (inverse Bremsstrahlung) \cite{Kaiser}. Second, electron excitation can also take place through a non-collisional process (no other particle as ion or phonon is involved to absorb photons) which is direct multiphoton interband transitions~\cite{J.PhysiqueIV.108.113, Belsky2004b, Belsky2004a, Yatsenko2005, Barilleau}. To include the latter process, the conduction band is described by multiple parabolic energy sub-bands. An illustration of this process is provided by Fig. \ref{fig:interband}. Note this mechanism departs from the Rethfeld's approach \cite{Rethfeld, Kaiser} and is expected to have a significant impact on the electron dynamics. The interband rate is evaluated according to the expression and parameters provided in \cite{Barilleau}. Such an approach allows us to introduce explicitly the collisionless heating in a full kinetic treatment of the electron dynamics in laser-driven dielectrics \cite{Barilleau}.

Solution to the Boltzmann equation (\ref{eq:Boltzmann}) provides the energy distribution of electrons in the material. Their ejection from the surface is possible if their energy is larger than the work function which is 0.9 eV for quartz \cite{JSIXSNT.5.764, SCT.202.5310}. For low energy ejection, surface effects may modify the distribution \cite{Conf.Proc.Italian.Phys.Soc.82.211}. However, this influence is negligible for the most energetic electrons which are considered here. Consequently, the distribution of ejected electrons near the surface is assumed to be the same as the one calculated in the bulk. To obtain a distribution \textit{directly} comparable to the experimental data, $f(E_k, t)$ is first weighted by the density of states $g(E_k)\propto \sqrt{E_k}$ accounting for a three-dimensional free electron gas. Secondly, the influence of the laser electric field $F(t)$, which may further accelerate or decelerate the ejected electrons depending on their instant of emission, is taken into account: it is a laser-driven field acceleration (LDFA) which can change the energy distribution \cite{NPHYS1983, NanoLett.13.674, Brunel}. The final energy of the ejected electron is obtained by integrating the classical equation of the electron motion in vacuum $dv/dt=-eF(t)/m_e$ from the ejection moment $t_e$ to the end of the laser pulse (160 fs in practice). The initial ejection velocity $v_0$ at $t_e$ is evaluated from the calculated electron distribution in the bulk. Since electrons in the bulk undergo numerous collisions before ejection, they lose any coherence \cite{Bourgeade} with the laser electric field at the time of ejection. Consequently, the electrons are assumed to be ejected uniformly during the interaction: their ejection time is not related to any particular phase of the laser electric field. Since in experiments ejected electrons are collected over the whole laser pulse duration, the theoretical predictions are obtained by integrating the electron distribution over time. Note that the maximum energy gain corresponds to the classical energy of \textit{half} an optical cycle. For an electron ejected at the optimal time, a simple calculation shows that the final energy is roughly 40 eV for an ejection energy $E_{k_0}=$ 20 eV and a laser intensity $I = 50 \ TW/cm^2$.
\begin{figure}[t]
  \includegraphics[width = 9cm, trim = 0cm 0cm 0cm 3cm, clip] {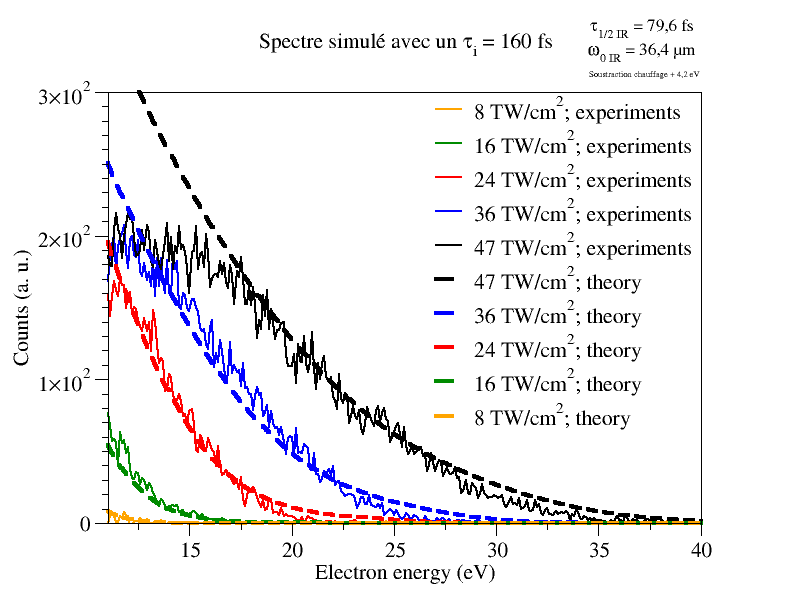}
  \caption{Energy distributions of ejected electrons from photo-emission experiments (solid lines) and modeling (dashed lines) for various laser intensities.}
  \label{fig:compDistrib}
\end{figure}

Figure \ref{fig:compDistrib} shows, within a linear scale, the experimental distributions of ejected electrons together with the theoretical predictions for intensities ranging from 8 to 47 TW/cm\textsuperscript{2}. The theoretical distribution at $I = 24$ TW/cm$^2$ is multiplied by a renormalization factor to compare to the experimental data. The same factor is used for all intensities. Both experimental and theoretical data are in a good agreement for each intensity. The only significant discrepancy between modeling and experimental spectra appears for electron energies below 17 eV for the largest laser intensity. It may be attributed to electron transport in the bulk which is not included in the present modeling: low energy electrons are sensitive to spatial charge rearrangements (potential minimization) in the bulk which are all the more important that the produced charge density is large (or highest laser intensities). 

The experimental observations can be reproduced only if we include all above-mentioned physical processes: photo-ionization, impact ionization, heating through electron-phonon-photon and interband transitions, the relaxation through electron-phonon and electron-electron collisions, and the LDFA. In order to evaluate the role of each process on the electron dynamics in the bulk, they have been successively switched off. The comparison of the theoretical spectra obtained with the various modeling configurations (not shown here) to the experimental data leads to the following conclusions. (i) the impact ionization prevents electrons from reaching too high energies at the largest intensities. However there is no electron avalanche, which is consistent with the fact that the irradiated material is not damaged. (ii) Regarding the electron heating in the conduction band, the introduction of both e-ph-pt and interband processes is required to recover correct slopes for all considered intensities. In particular, the interband process enables to mimic the smooth decrease with respect to the electron energy \cite{Barilleau}. (iii) Regarding the relaxation, the electron-electron collisions provide a smooth energy distribution. Otherwise the electron distribution contains several peaks separated by the photon energy \cite{Barilleau} that is not experimentally observed. The contribution of e-ph collisions also permits to redistribute electrons to lower energies providing the observed slopes. In contrast, the electron recombination and e-ion-pt do not modify significantly the spectra due to the short interaction time and relatively low ionization degree (the electron density in the CB is in between $10^{19}$ and $10^{20}$ cm$^{-3}$ depending on the intensity), respectively.

% Phrase a introduire??????
% relative completeness of the modeling : include all possible physical processes

%{\notedebase{This difference may be attributed to the formation of defects which may introduce states located in the band gap. Slow electrons may be trapped easily in these defective states leading to a decrease in the slow energy population as experimentally observed. Furthermore, this discrepancy may also be due to the electronic transport which is not included in the present modeling. \textbf{Devrait être observé aussi à bas flux dans le modèle!}}}

%{\notedebase{In particular, the Fermi-Dirac distribution shape is well reproduced. We conclude that the energy distribution of electrons in the conduction band is close to an equilibrium distribution before the end of the laser pulse. This result was not expected since, in general, it is assumed that energetic electrons are ejected without undergoing inelastic collisions. It follows that all the above-mentioned physical processes of interactions (collisions and LDFA) contribute to the electron dynamics whatever its energy. Furthermore, the higher the intensity, the smoother the part related to the Fermi-Dirac distribution. This behavior accounts for an increase in the average energy of the electron gas which is of the order of a few eV. \textbf{A voir avec la distribution des électrons dans le bulk!}}}

\begin{figure}[t]
  \includegraphics[width = 9cm, trim = 0cm 0cm 0cm 3cm, clip] {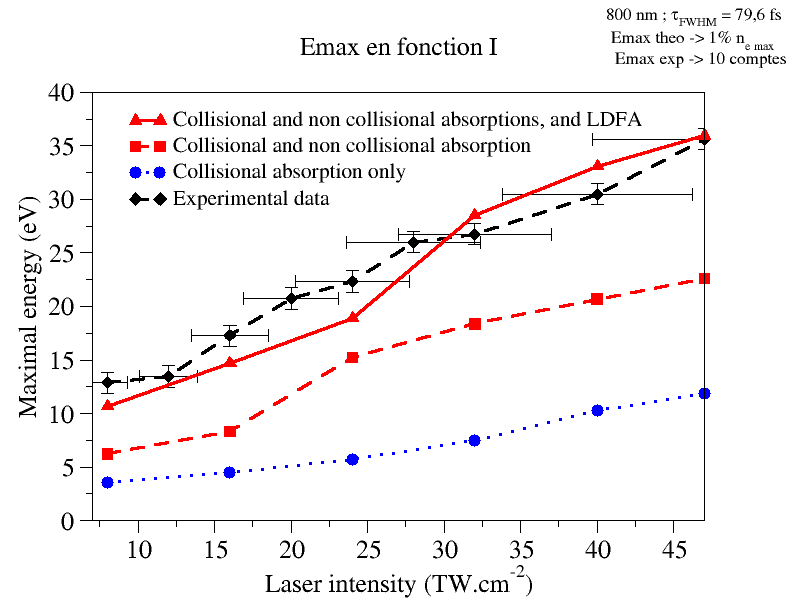}
  \caption{Maximum energy of photoemitted electrons as a function of the laser intensity. Absorption processes are gradually switched on within the modeling, see inset legend for curves meaning.}
  \label{image_E_max}
\end{figure}
Figure \ref{image_E_max} provides the evolution of $E_{max}$ as a function of the laser intensity from experimental observations and as predicted by the modeling (the interband and LDFA processes are included or not). In the experiment, $E_{max}$ increases monotonically from 11 eV to roughly 40 eV. Without the interband and LDFA processes, $E_{max}$ cannot exceed 10 eV for the highest intensity, the heating being only due to e-ph-pt collisions in that case. By including the interband process, $E_{max}$ reaches about 23 eV, i.e. twice the energy of the previous configuration. Both e-ph-pt and interband contributions to the electron heating are thus comparable. A good agreement with the experimental data is obtained when the LDFA is included, providing an enhancement of the final electron energy of more than a factor of 2. These considerations clearly show that the observed photo-emission spectra result from three physical processes with comparable contributions. All the previous conclusions are expected to be similar for the above-mentioned other large band gap dielectric materials. A quantitative reproduction of other data being able by slightly adjusting a few modeling parameters as the value of the band gap and matrix elements.

\begin{figure}[t]
  \includegraphics[width = 9cm, clip] {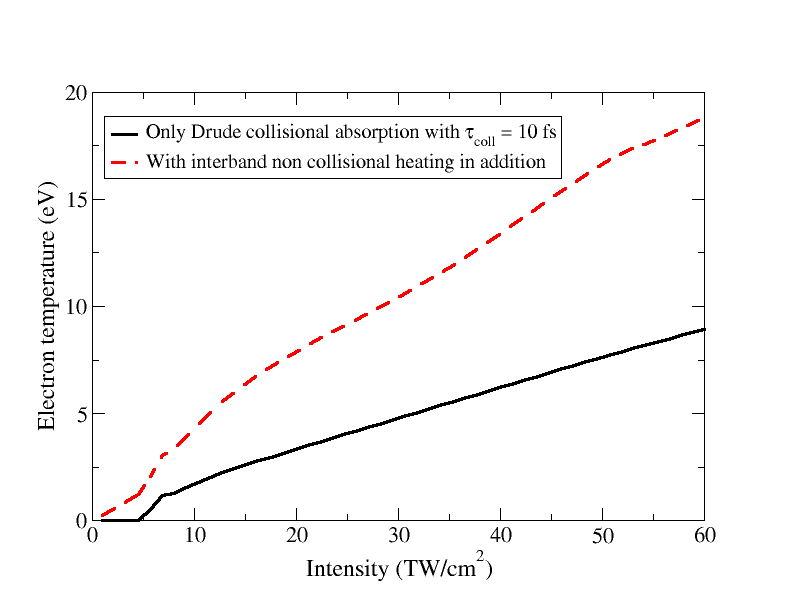}
  \caption{Evolution of the electron temperature as a function of the maximum laser intensity as estimated with a simplified Drude-like modeling.}
  \label{temperatures}
\end{figure}

% Discussion plus generale, Drude devrait etre revisite car pas juste processus collisionel.
We have shown the importance of the non collisional laser heating. For applicative purposes as laser structuring of materials, a simple expression to evaluate the laser energy deposition is desirable. It can be shown that the energy density absorbed per unit of time through the interband process, $dU_{nc}/dt$, reads:
\begin{equation}
  \frac{dU_{nc}}{dt}  = \frac{m\omega n_e(t)(\Delta k / (\pi/a))}{4 \pi \hbar \left| p_f \right|} V_{1f}^2 I \sum\limits_{n} n J'^2_n(B_{1f}) \left[ \left( \frac{\pi}{a} \right)^2 - k_n^2 \right]
  \label{Wif}
\end{equation}
%Considering applications as laser material structuring, an accurate estimation of the laser energy deposition into the material is required \cite{Mazur, toto}. For that purpose, the Drude model based on collisional absorption is widely used; it predicts a linear energy deposition with respect to the laser intensity. The energy density absorbed per unit of time, $dU_{c}/dt$, reads:
%\begin{equation}
%  \frac{dU_{c}}{dt} = \frac{2 e^2 n_{fe} \nu_c}{m_e c \varepsilon_0 (\omega^2 + \nu_c^2)}I
%\end{equation}
%We have shown this collisional absorption has to be corrected by the non collisional interband process. This process is also expected to play a major role for the impact ionization and criteria for electron avalanche to engage. Based on the expression of the interband rate \cite{Barilleau}, 
%
%
%a simple cubic lattice structure is considered with a period of 0.49 nm \cite{Barilleau}. The interband heating rate depends on a dipolar matrix element bridging conduction sub-bands. It has been set to $V_{1f} = 9\times 10^{-55}$ J.m\textsuperscript{2}.s according to an evaluation based on the numerical solution of the Schrödinger equation \cite{PhysRevB.74.235215, Europhys.Lett.67.301, Barilleau}. 
%
with $B_{1f} = \frac{1}{\hbar \omega} \frac{e \vec{F} \left( t \right) . \left( \vec{p}_f - \vec{p}_1 \right)}{m \omega}$. All notations and values of parameters relative to the interband rate are defined in \cite{Barilleau}. $n_e(t)$ is the electron density in the conduction band which here is evaluated by solving multiple rate equations \cite{Rethfeld}. $\Delta k / (\pi/a) = 2 m a^2 h \nu_c / \pi^2 \hbar^2$ is the relative part of the Brillouin zone participating to interband transition due to the collisional broadening (see Fig. \ref{fig:interband}). By setting the Drude averaged collision time to $\nu_c^{-1} =$ 10 fs accounting mainly for electron-phonon collisions, the evolutions of the electron temperature ($= U/C_e$ with $C_e$ the classical heat capacity) as a function of the intensity including or not the non collisional laser heating are obtained (see Fig. \ref{temperatures}). They exhibit similar trends as those provided by solving the quantum Boltzmann equation \cite{Barilleau}, and values consistent with the present electron energies (because $T_e \simeq E_{max} / 2$ \cite{Barilleau}). This demonstrates the reliability of this simplified model (\ref{Wif}) describing the additional contribution of non collisional heating to the Drude description.

%%%%%%%%%%%%%%%%%%%%%%%%%%%%%%%%%%%%%%%%%%%%%%%%%%%%%%%%%%%%%%%%%%%%%%%%%%%%%%%%%%%%%%%%%%%%%%%%%
%%%%%%%%%%% CONCLUSION
%%%%%%%%%%%%%%%%%%%%%%%%%%%%%%%%%%%%%%%%%%%%%%%%%%%%%%%%%%%%%%%%%%%%%%%%%%%%%%%%%%%%%%%%%%%%%%%%%
To conclude, photo-emission experiments have been carried out with large band gap dielectric crystals irradiated by near infrared laser femtosecond pulses with intensities below the ablation threshold. The electron energy spectra exhibit a long tail up to energies close to 40 eV for the highest intensities. The underlying electron dynamics has been analyzed through a state-of-the-art modeling based on the Boltzmann kinetic equation including the main excitation/relaxation processes, and the laser driven field acceleration of ejected electrons. A direct comparison of the theoretical predictions to the experimental data shows that both heating in the bulk and electric field acceleration in the vacuum make comparable contributions to the electron energy gain. The noncollisional direct multiphoton transitions between sub-bands of the conduction band is a major mechanism for electron heating in the bulk of dielectric materials which must be included for the evaluation of the energy deposition. For application purpose as laser structuring of materials, a simple expression to evaluate the energy deposition by non collisional absorption can be derived \cite{suppMat}.

\begin{acknowledgments}
  This project has received funding from the European Union’s Horizon 2020 research and innovation programme under grant agreement No 654148 Laserlab-Europe.
\end{acknowledgments}

%\bibliography{A-Bibliographie}

\begin{thebibliography}{30}%
\makeatletter
\providecommand \@ifxundefined [1]{%
 \@ifx{#1\undefined}
}%
\providecommand \@ifnum [1]{%
 \ifnum #1\expandafter \@firstoftwo
 \else \expandafter \@secondoftwo
 \fi
}%
\providecommand \@ifx [1]{%
 \ifx #1\expandafter \@firstoftwo
 \else \expandafter \@secondoftwo
 \fi
}%
\providecommand \natexlab [1]{#1}%
\providecommand \enquote  [1]{``#1''}%
\providecommand \bibnamefont  [1]{#1}%
\providecommand \bibfnamefont [1]{#1}%
\providecommand \citenamefont [1]{#1}%
\providecommand \href@noop [0]{\@secondoftwo}%
\providecommand \href [0]{\begingroup \@sanitize@url \@href}%
\providecommand \@href[1]{\@@startlink{#1}\@@href}%
\providecommand \@@href[1]{\endgroup#1\@@endlink}%
\providecommand \@sanitize@url [0]{\catcode `\\12\catcode `\$12\catcode
  `\&12\catcode `\#12\catcode `\^12\catcode `\_12\catcode `\%12\relax}%
\providecommand \@@startlink[1]{}%
\providecommand \@@endlink[0]{}%
\providecommand \url  [0]{\begingroup\@sanitize@url \@url }%
\providecommand \@url [1]{\endgroup\@href {#1}{\urlprefix }}%
\providecommand \urlprefix  [0]{URL }%
\providecommand \Eprint [0]{\href }%
\providecommand \doibase [0]{http://dx.doi.org/}%
\providecommand \selectlanguage [0]{\@gobble}%
\providecommand \bibinfo  [0]{\@secondoftwo}%
\providecommand \bibfield  [0]{\@secondoftwo}%
\providecommand \translation [1]{[#1]}%
\providecommand \BibitemOpen [0]{}%
\providecommand \bibitemStop [0]{}%
\providecommand \bibitemNoStop [0]{.\EOS\space}%
\providecommand \EOS [0]{\spacefactor3000\relax}%
\providecommand \BibitemShut  [1]{\csname bibitem#1\endcsname}%
\let\auto@bib@innerbib\@empty
%</preamble>
\bibitem [{\citenamefont {RETHFELD}\ \emph {et~al.}(2002)\citenamefont
  {RETHFELD}, \citenamefont {KAISER}, \citenamefont {VICANEK},\ and\
  \citenamefont {SIMON}}]{Rethfeldb}%
  \BibitemOpen
  \bibfield  {author} {B.~Rethfeld, A.~Kaiser, M.~Vicanek, \ and\ G.~Simon,\
  }\href@noop {} {\bibfield  {journal} {\bibinfo  {journal} {Phys. Rev. B}\
  }\textbf {\bibinfo {volume} {65}},\ \bibinfo {pages} {214303(1)} (\bibinfo
  {year} {2002})}\BibitemShut {NoStop}%
\bibitem [{\citenamefont {SCHARF}\ \emph {et~al.}(2013)\citenamefont {SCHARF},
  \citenamefont {PEREBEINOS}, \citenamefont {FABIAN},\ and\ \citenamefont
  {AVOURIS}}]{PhysRevB.87.035414}%
  \BibitemOpen
  \bibfield  {author} {B.~Scharf, V.~Perebeinos, J.~Fabian, \ and\ P.~Avouris,\
  }\href@noop {} {\bibfield  {journal} {\bibinfo  {journal} {Phys. Rev. B}\
  }\textbf {\bibinfo {volume} {87}},\ \bibinfo {pages} {035414(1)} (\bibinfo
  {year} {2013})}\BibitemShut {NoStop}%
\bibitem [{\citenamefont {SELBMANN}\ \emph {et~al.}(1996)\citenamefont
  {SELBMANN}, \citenamefont {GULIA}, \citenamefont {ROSSI},\ and\ \citenamefont
  {MOLINAR}}]{PhysRevB.54.4660}%
  \BibitemOpen
  \bibfield  {author} {P.~E. Selbmann, M.~Gulia, F.~Rossi, \ and\ E.~Molinar,\
  }\href@noop {} {\bibfield  {journal} {\bibinfo  {journal} {Phys. Rev. B}\
  }\textbf {\bibinfo {volume} {54}},\ \bibinfo {pages} {4660} (\bibinfo {year}
  {1996})}\BibitemShut {NoStop}%
\bibitem [{\citenamefont {KAISER}\ \emph {et~al.}(2000)\citenamefont {KAISER},
  \citenamefont {RETHEFELD}, \citenamefont {VICANEK},\ and\ \citenamefont
  {SIMON}}]{Kaiser}%
  \BibitemOpen
  \bibfield  {author} {A.~Kaiser, B.~Rethefeld, M.~Vicanek, \ and\ G.~Simon,\
  }\href@noop {} {\bibfield  {journal} {\bibinfo  {journal} {Phys. Rev. B}\
  }\textbf {\bibinfo {volume} {61}},\ \bibinfo {pages} {11437} (\bibinfo {year}
  {2000})}\BibitemShut {NoStop}%
\bibitem [{\citenamefont {Shcheblanov}\ and\ \citenamefont
  {Itina}(2013)}]{Shchelebanov}%
  \BibitemOpen
  \bibfield  {author} {N.~S. Shcheblanov\ and\ T.~E. Itina,\ }\href@noop {}
  {\bibfield  {journal} {\bibinfo  {journal} {Appl. Phys. A}\ }\textbf
  {\bibinfo {volume} {110}},\ \bibinfo {pages} {579} (\bibinfo {year}
  {2013})}\BibitemShut {NoStop}%
\bibitem [{\citenamefont {Glezer}\ \emph {et~al.}(1996)\citenamefont {Glezer},
  \citenamefont {Milosavljevic}, \citenamefont {Huang}, \citenamefont {Finlay},
  \citenamefont {Her}, \citenamefont {Callan},\ and\ \citenamefont
  {Mazur}}]{Glezer96}%
  \BibitemOpen
  \bibfield  {author} {E.~N. Glezer, M.~Milosavljevic, L.~Huang, R.~J. Finlay,
  T.-H. Her, J.~P. Callan, \ and\ E.~Mazur,\ }\href {\doibase
  10.1364/OL.21.002023} {\bibfield  {journal} {\bibinfo  {journal} {Opt.
  Lett.}\ }\textbf {\bibinfo {volume} {21}},\ \bibinfo {pages} {2023} (\bibinfo
  {year} {1996})}\BibitemShut {NoStop}%
\bibitem [{\citenamefont {Gamaly}()}]{Gamaly}%
  \BibitemOpen
  \bibfield  {author} {E.G. Gamaly,\ }\href@noop {} {\bibfield  {journal}
  {\bibinfo  {journal} {Physics Reports}\ }\textbf {\bibinfo {volume} {508}},\
  \bibinfo {pages} {91}}\BibitemShut {NoStop}%
\bibitem [{\citenamefont {Marcinkevi\v{c}ius}\ \emph
  {et~al.}(2001)\citenamefont {Marcinkevi\v{c}ius}, \citenamefont {Juodkazis},
  \citenamefont {Watanabe}, \citenamefont {Miwa}, \citenamefont {Matsuo},
  \citenamefont {Misawa},\ and\ \citenamefont {Nishii}}]{Juodkazis}%
  \BibitemOpen
  \bibfield  {author} {Andrius Marcinkevi\v{c}ius, Saulius Juodkazis, Mitsuru
  Watanabe, Masafumi Miwa, Shigeki Matsuo, Hiroaki Misawa, \ and\ Junji
  Nishii,\ }\href {\doibase 10.1364/OL.26.000277} {\bibfield  {journal}
  {\bibinfo  {journal} {Opt. Lett.}\ }\textbf {\bibinfo {volume} {26}},\
  \bibinfo {pages} {277} (\bibinfo {year} {2001})}\BibitemShut {NoStop}%
\bibitem [{\citenamefont {Gattass}\ and\ \citenamefont {Mazur}(2008)}]{Mazur}%
  \BibitemOpen
  \bibfield  {author} {Rafael~R. Gattass\ and\ Eric Mazur,\ }\href@noop {}
  {\bibfield  {journal} {\bibinfo  {journal} {Nature Photonics}\ }\textbf
  {\bibinfo {volume} {2}},\ \bibinfo {pages} {219–225} (\bibinfo {year}
  {2008})}\BibitemShut {NoStop}%
\bibitem [{\citenamefont {ZHANG}\ \emph {et~al.}(2014)\citenamefont {ZHANG},
  \citenamefont {GECEVICIUS}, \citenamefont {BERESNA},\ and\ \citenamefont
  {KAZANSKY}}]{PhysRevLett.112.033901}%
  \BibitemOpen
  \bibfield  {author} {J.~Zhang, M.~Gecevicius, M.~Beresna, \ and\ P.~G.
  Kazansky,\ }\href@noop {} {\bibfield  {journal} {\bibinfo  {journal} {Phys.
  Rev. Lett.}\ }\textbf {\bibinfo {volume} {112}},\ \bibinfo {pages} {033901}
  (\bibinfo {year} {2014})}\BibitemShut {NoStop}%
\bibitem [{\citenamefont {Menzel}(2001)}]{Menzel}%
  \BibitemOpen
  \bibfield  {author} {Ralf Menzel,\ }\href@noop {} {\emph {\bibinfo {title}
  {Photonics: Linear and Nonlinear Interactions of Laser Light and Matter}}}\
  (\bibinfo  {publisher} {Springer},\ \bibinfo {year} {2001})\BibitemShut
  {NoStop}%
\bibitem [{\citenamefont {RETHFELD}(2004)}]{Rethfeld}%
  \BibitemOpen
  \bibfield  {author} {B.~Rethfeld,\ }\href@noop {} {\bibfield  {journal}
  {\bibinfo  {journal} {Phys. Rev. Lett.}\ }\textbf {\bibinfo {volume} {92}},\
  \bibinfo {pages} {187401} (\bibinfo {year} {2004})}\BibitemShut {NoStop}%
\bibitem [{\citenamefont {Barilleau}\ \emph {et~al.}(2016)\citenamefont
  {Barilleau}, \citenamefont {Duchateau}, \citenamefont {Chimier},
  \citenamefont {Geoffroy},\ and\ \citenamefont {Tikhonchuk}}]{Barilleau}%
  \BibitemOpen
  \bibfield  {author} {L.~Barilleau, G.~Duchateau, B.~Chimier, G.~Geoffroy, \
  and\ V.~Tikhonchuk,\ }\href@noop {} {\bibfield  {journal} {\bibinfo
  {journal} {Journal of Physics D: Applied Physics}\ }\textbf {\bibinfo
  {volume} {49}},\ \bibinfo {pages} {485103} (\bibinfo {year}
  {2016})}\BibitemShut {NoStop}%
\bibitem [{\citenamefont {DOMBI}\ \emph {et~al.}(2013)\citenamefont {DOMBI},
  \citenamefont {HORLAND}, \citenamefont {RACZ}, \citenamefont {MARTON},
  \citenamefont {TRUGLER}, \citenamefont {KRENN},\ and\ \citenamefont
  {HOHENESTER}}]{NanoLett.13.674}%
  \BibitemOpen
  \bibfield  {author} {P.~Dombi, A.~Horland, P.~Racz, I.~Marton, A.~Trugler,
  J.~R. Krenn, \ and\ U.~Hohenester,\ }\href@noop {} {\bibfield  {journal}
  {\bibinfo  {journal} {Nano Lett.}\ }\textbf {\bibinfo {volume} {13}},\
  \bibinfo {pages} {674} (\bibinfo {year} {2013})}\BibitemShut {NoStop}%
\bibitem [{\citenamefont {YATSENKO}\ \emph {et~al.}(2005)\citenamefont
  {YATSENKO}, \citenamefont {BACHAU}, \citenamefont {BELSKY}, \citenamefont
  {GAUDIN}, \citenamefont {GEOFFROY}, \citenamefont {GUIZARD}, \citenamefont
  {MARTIN}, \citenamefont {PETITE}, \citenamefont {PHILIPPOV},\ and\
  \citenamefont {VASIL'EV}}]{Yatsenko2005}%
  \BibitemOpen
  \bibfield  {author} {B.~N. Yatsenko, H.~Bachau, A.~Belsky, J.~Gaudin,
  G.~Geoffroy, S.~Guizard, P.~Martin, G.~Petite, A.~Philippov, \ and\ A.~N.
  Vasil'ev,\ }\href@noop {} {\bibfield  {journal} {\bibinfo  {journal} {Phys.
  Status. Solidi. C}\ }\textbf {\bibinfo {volume} {2}},\ \bibinfo {pages} {240}
  (\bibinfo {year} {2005})}\BibitemShut {NoStop}%
\bibitem [{\citenamefont {BACHAU}\ \emph {et~al.}(2009)\citenamefont {BACHAU},
  \citenamefont {BELSKY}, \citenamefont {BOGATYREV}, \citenamefont {GAUDIN},
  \citenamefont {GEOFFROY}, \citenamefont {GUIZARD}, \citenamefont {MARTIN},
  \citenamefont {POPOV}, \citenamefont {VASIL{'}EV},\ and\ \citenamefont
  {YATSENKO}}]{Bachau2010}%
  \BibitemOpen
  \bibfield  {author} {H.~Bachau, A.~N. Belsky, I.~B. Bogatyrev, J.~Gaudin,
  G.~Geoffroy, S.~Guizard, P.~Martin, Yu.~V. Popov, A.~N. Vasil{'}ev, \ and\
  B.~N. Yatsenko,\ }\href@noop {} {\bibfield  {journal} {\bibinfo  {journal}
  {Appl. Phys. A}\ }\textbf {\bibinfo {volume} {98}},\ \bibinfo {pages} {679}
  (\bibinfo {year} {2009})}\BibitemShut {NoStop}%
\bibitem [{\citenamefont {BELSKY}\ \emph
  {et~al.}(2004{\natexlab{a}})\citenamefont {BELSKY}, \citenamefont {BACHAU},
  \citenamefont {GAUDIN}, \citenamefont {GEOFFROY}, \citenamefont {GUIZARD},
  \citenamefont {MARTIN}, \citenamefont {PETITE}, \citenamefont {PHILIPPOV},
  \citenamefont {VASIL'EV},\ and\ \citenamefont {YATSENKO}}]{Belsky2004a}%
  \BibitemOpen
  \bibfield  {author} {A.~Belsky, H.~Bachau, J.~Gaudin, G.~Geoffroy,
  S.~Guizard, P.~Martin, G.~Petite, A.~Philippov, A.~N. Vasil'ev, \ and\ B.~N.
  Yatsenko,\ }\href@noop {} {\bibfield  {journal} {\bibinfo  {journal} {Appl.
  Phys. B}\ }\textbf {\bibinfo {volume} {78}},\ \bibinfo {pages} {989}
  (\bibinfo {year} {2004}{\natexlab{a}})}\BibitemShut {NoStop}%
\bibitem [{\citenamefont {BELSKY}\ \emph
  {et~al.}(2004{\natexlab{b}})\citenamefont {BELSKY}, \citenamefont {MARTIN},
  \citenamefont {BACHAU}, \citenamefont {VASIL'EV}, \citenamefont {YATSENKO},
  \citenamefont {GUIZARD}, \citenamefont {GEOFFROY},\ and\ \citenamefont
  {PETITE}}]{Belsky2004b}%
  \BibitemOpen
  \bibfield  {author} {A.~Belsky, P.~Martin, H.~Bachau, A.~N. Vasil'ev, B.~N.
  Yatsenko, S.~Guizard, G.~Geoffroy, \ and\ G.~Petite,\ }\href@noop {}
  {\bibfield  {journal} {\bibinfo  {journal} {Europhys. Lett.}\ }\textbf
  {\bibinfo {volume} {67}},\ \bibinfo {pages} {301} (\bibinfo {year}
  {2004}{\natexlab{b}})}\BibitemShut {NoStop}%
\bibitem [{\citenamefont {ZHEREBTSOV}\ \emph {et~al.}(2011)\citenamefont
  {ZHEREBTSOV}, \citenamefont {FENNEL}, \citenamefont {PLENGE}, \citenamefont
  {ANTONSSON}, \citenamefont {ZNAKOVSKAYA}, \citenamefont {WIRTH},
  \citenamefont {HEREWERTH}, \citenamefont {SUBMANN}, \citenamefont {PELTZ},
  \citenamefont {AHMAD}, \citenamefont {TRUSHIN}, \citenamefont {PERVAK},
  \citenamefont {KARSCH}, \citenamefont {VRAKKING}, \citenamefont {LANGER},
  \citenamefont {GRAF}, \citenamefont {STOKMAN}, \citenamefont {KRAUSZ},
  \citenamefont {RUHL},\ and\ \citenamefont {KLING}}]{NPHYS1983}%
  \BibitemOpen
  \bibfield  {author} {S.~Zherebtsov, T.~Fennel, J.~Plenge, E.~Antonsson,
  I.~Znakovskaya, A.~Wirth, O.~Herewerth, F.~Submann, C.~Peltz, I.~Ahmad, S.~A.
  Trushin, V.~Pervak, S.~Karsch, M.~J.~J. Vrakking, B.~Langer, C.~Graf, M.~I.
  Stokman, F.~Krausz, E.~Ruhl, \ and\ M.~F. Kling,\ }\href@noop {} {\bibfield
  {journal} {\bibinfo  {journal} {N. Phys.}\ }\textbf {\bibinfo {volume} {7}},\
  \bibinfo {pages} {656} (\bibinfo {year} {2011})}\BibitemShut {NoStop}%
\bibitem [{\citenamefont {BAGNOUD}\ and\ \citenamefont
  {SALIN}(2000)}]{Appl.Phys.B.70.S165}%
  \BibitemOpen
  \bibfield  {author} {V.~Bagnoud\ and\ F.~Salin,\ }\href@noop {} {\bibfield
  {journal} {\bibinfo  {journal} {Appl. Phys. B}\ }\textbf {\bibinfo {volume}
  {70}},\ \bibinfo {pages} {S165} (\bibinfo {year} {2000})}\BibitemShut
  {NoStop}%
\bibitem [{\citenamefont {DAGUZAN}\ \emph {et~al.}(1995)\citenamefont
  {DAGUZAN}, \citenamefont {MARTIN}, \citenamefont {GUIZARD},\ and\
  \citenamefont {PETITE}}]{Daguzan}%
  \BibitemOpen
  \bibfield  {author} {Ph. Daguzan, P.~Martin, S.~Guizard, \ and\ G.~Petite,\
  }\href@noop {} {\bibfield  {journal} {\bibinfo  {journal} {Phys. Rev. B}\
  }\textbf {\bibinfo {volume} {52}},\ \bibinfo {pages} {17099} (\bibinfo {year}
  {1995})}\BibitemShut {NoStop}%
\bibitem [{\citenamefont {KELDYSH}(1965)}]{Keldysh}%
  \BibitemOpen
  \bibfield  {author} {L.~V. Keldysh,\ }\href@noop {} {\bibfield  {journal}
  {\bibinfo  {journal} {Sov. Phys. JETP}\ }\textbf {\bibinfo {volume} {20}},\
  \bibinfo {pages} {1307} (\bibinfo {year} {1965})}\BibitemShut {NoStop}%
\bibitem [{\citenamefont {MARTIN}\ \emph {et~al.}(1997)\citenamefont {MARTIN},
  \citenamefont {GUIZARD}, \citenamefont {DAGUZAN},\ and\ \citenamefont
  {PETITE}}]{Martin}%
  \BibitemOpen
  \bibfield  {author} {P.~Martin, S.~Guizard, Ph. Daguzan, \ and\ G.~Petite,\
  }\href@noop {} {\bibfield  {journal} {\bibinfo  {journal} {Phys. Rev. B}\
  }\textbf {\bibinfo {volume} {55}},\ \bibinfo {pages} {5799} (\bibinfo {year}
  {1997})}\BibitemShut {NoStop}%
\bibitem [{sup()}]{suppMat}%
  \BibitemOpen
  \href@noop {} {\ }\bibinfo {note} {See Supplemental Material at [URL will be
  inserted by publisher] for an averaged expression of the non collisional
  laser absorption.}\BibitemShut {Stop}%
\bibitem [{\citenamefont {BELSKY}\ \emph {et~al.}(2003)\citenamefont {BELSKY},
  \citenamefont {VASIL'EV}, \citenamefont {YATSENKO}, \citenamefont {BACHAU},
  \citenamefont {MARTIN}, \citenamefont {GEOFFROY},\ and\ \citenamefont
  {GUIZARD}}]{J.PhysiqueIV.108.113}%
  \BibitemOpen
  \bibfield  {author} {A.~Belsky, A.~N. Vasil'ev, B.~N. Yatsenko, H.~Bachau,
  P.~Martin, G.~Geoffroy, \ and\ S.~Guizard,\ }\href@noop {} {\bibfield
  {journal} {\bibinfo  {journal} {J. Phys. IV}\ }\textbf {\bibinfo {volume}
  {108}},\ \bibinfo {pages} {113} (\bibinfo {year} {2003})}\BibitemShut
  {NoStop}%
\bibitem [{\citenamefont {KORTOV}\ and\ \citenamefont
  {ZVONAREV}(2011)}]{JSIXSNT.5.764}%
  \BibitemOpen
  \bibfield  {author} {V.~S. Kortov\ and\ S.~V. Zvonarev,\ }\href@noop {}
  {\bibfield  {journal} {\bibinfo  {journal} {Journal of Surface Investigation.
  X-ray, Synchrotron and Neutron Techniques}\ }\textbf {\bibinfo {volume}
  {5}},\ \bibinfo {pages} {764} (\bibinfo {year} {2011})}\BibitemShut {NoStop}%
\bibitem [{\citenamefont {OHYA}\ \emph {et~al.}(2008)\citenamefont {OHYA},
  \citenamefont {INAI}, \citenamefont {KUWADA}, \citenamefont {HAYASHI},\ and\
  \citenamefont {SAITO}}]{SCT.202.5310}%
  \BibitemOpen
  \bibfield  {author} {K.~Ohya, K.~Inai, H.~Kuwada, T.~Hayashi, \ and\
  M.~Saito,\ }\href@noop {} {\bibfield  {journal} {\bibinfo  {journal} {Surface
  and Coatings Technology}\ }\textbf {\bibinfo {volume} {202}},\ \bibinfo
  {pages} {5310} (\bibinfo {year} {2008})}\BibitemShut {NoStop}%
\bibitem [{\citenamefont {MARIANI}(2003)}]{Conf.Proc.Italian.Phys.Soc.82.211}%
  \BibitemOpen
  \bibfield  {author} {C.~Mariani,\ }\href@noop {} {\bibfield  {journal}
  {\bibinfo  {journal} {Conf. Proc. Italian Phys. Soc.}\ }\textbf {\bibinfo
  {volume} {82}},\ \bibinfo {pages} {211} (\bibinfo {year} {2003})}\BibitemShut
  {NoStop}%
\bibitem [{\citenamefont {Mulser}\ \emph {et~al.}(2012)\citenamefont {Mulser},
  \citenamefont {Weng},\ and\ \citenamefont {Liseykina}}]{Brunel}%
  \BibitemOpen
  \bibfield  {author} {P.~Mulser, S.~M. Weng, \ and\ Tatyana Liseykina,\
  }\href@noop {} {\bibfield  {journal} {\bibinfo  {journal} {Phys. Plasmas}\
  }\textbf {\bibinfo {volume} {19}},\ \bibinfo {pages} {043301} (\bibinfo
  {year} {2012})}\BibitemShut {NoStop}%
\bibitem [{\citenamefont {BOURGEADE}\ and\ \citenamefont
  {DUCHATEAU}(2012)}]{Bourgeade}%
  \BibitemOpen
  \bibfield  {author} {A.~Bourgeade\ and\ G.~Duchateau,\ }\href@noop {}
  {\bibfield  {journal} {\bibinfo  {journal} {Phys. Rev. E}\ }\textbf {\bibinfo
  {volume} {85}},\ \bibinfo {pages} {056403} (\bibinfo {year}
  {2012})}\BibitemShut {NoStop}%
\end{thebibliography}
%\bibliographystyle{apsrev4-1}

%

\end{document}